\documentclass[a4paper,11pt]{article}
\usepackage{pos}

\newcommand{\pTf}{p_{\scriptscriptstyle T}}
\newcommand{\pTi}{p_{\scriptscriptstyle T}^{\rm \tiny initial}}

\newcommand{\be}{\begin{equation}}
\newcommand{\ee}{\end{equation}}

\title{Overview: Jet quenching with machine learning}

\author*{Yi-Lun Du}

\affiliation{Shandong Institute of Advanced Technology, Jinan 250100, China}

\emailAdd{yilun.du@iat.cn}

\abstract{Jets are suppressed and modified in heavy ion collisions, which serve as powerful probes to the properties of the quark-gluon plasma (QGP). Attributed to the abundant information carried by the jet constituents and reconstructed substructures, plenty of interesting applications of machine learning techniques have been made on a jet-by-jet basis to study the jet quenching phenomena. Here we review recent proceedings on this topic including the tasks of reconstructing jet momentum in heavy ion collisions, classifying quenched jets and unquenched jets, identifying jet energy loss, locating the jet creation points as well as distinguishing between quark- and gluon-initiated jets in the QGP. Such jet-by-jet analyses will allow us to have a better handle on the jet reconstruction and selections to investigate the effects of jet modifications and push forward the long-standing goal of jet tomographic probes of the QGP.}

\FullConference{HardProbes2023\\
 26-31 March 2023\\
 Aschaffenburg, Germany\\}

\begin{document}
\maketitle

\section{Introduction}
QCD predicts that nuclear matter will form a new state of matter, i.e., quark-gluon plasma (QGP) at high temperature and density where quarks and gluons are deconfined from hadronic matter to form a strongly-coupled viscous fluid. The experiments of relativistic heavy ion collisions are conducted at RHIC and LHC to explore the nature of the new state of matter. In high-energy particle collisions, jets are collimated sprays of hadrons generated in a hard QCD process. While in heavy ion collisions, energetic partons will lose energy via the interactions with the QGP during their passage. The lost energy will hadronize and be redistributed around the energetic partons. These processes will quench the jet energy and modify the jet substructures. Besides, the interactions of QGP with high-energy partons will also generate the excitation or response of the medium. Mach cones are also expected to form in the expanding QGP when the energetic partons traverse the hot medium at a velocity faster than the speed of sound. Eventually some particles from the medium will stay inside the jets, which poses a challenge to the background subtraction of jets in heavy ion collisions. With such interesting interplay between jets and the QGP, high-energy hadrons or jets are employed as unique probes to the properties of the QGP~\cite{Mehtar-Tani:2013pia,Qin:2015srf}. Jet quenching is clearly manifest when calculating the ratio of the yields of high-energy hadrons or jets between those measured in heavy ion collisions and proton-proton collisions~\cite{adcox2001suppression,adler2002centrality,Abelev:2013kqa}. 

Additionally, one can also study the medium modifications of jets by analyzing their substructures, again usually done by comparing the results of jets measured in nucleus-nucleus collisions against those measured in proton-proton collisions. 
However, when comparing the quenched jets and unquenched jets at the same final, measured energy range, one needs to take into account the presence of a selection bias. Due to the steeply falling jet spectrum, jets losing too much energy will be under-represented after imposing the selection criteria. In other words, the selected, surviving jet samples generally possess the characteristics of the jet substructures that tend to lose less energy, hindering in this way our interpretation about what true medium-induced modifications of jet substructures are. These ambiguities affecting a typical analysis could be mitigated if one can estimate the jet energy loss on a jet-by-jet basis, which allow us to classify and select them according to their degree of modification. 

Towards a tomographic study of the QGP using jets, more jet-by-jet analyses are highly requested in many aspects to tackle the complicated jet-medium interactions. On the one hand, the QGP could have different fluctuating spatio-temporal profiles of temperature and density due to the relativistic collisions of different nucleus with different collision energy and overlapping geometry. On the other hand, jets could be initiated with energetic partons of different flavors, (transverse) momentum and energy at different positions inside the QGP. Then they could develop substructures in the early stage before the medium effects jump in and traverse different lengths through the regions of different temperature inside the QGP according to their orientations. If one can effectively pin down these early-stage uncertainties of jet partons before the quenching, the capabilities of jets for a tomographic study of the QGP will be enhanced to an unprecedented level.

In recent years, machine learning, especially deep learning techniques, have shown powerful capabilities on the data analysis. These novel techniques are good at digging out hidden correlations from big data, achieving widespread success in physics~\cite{mott2017solving,carrasquilla2017machine,van2017learning,pang2018equation,zhou2019regressive,du2020identifying,LonggangPANG2020Deep}. In particular, attributed to the rich information carried by the jet constituents, machine learning has been widely applied in the studies of jets in high-energy physics~\cite{feickert2021living}, such as QCD/W jet tagging, top jet tagging, quark/gluon jet classification and heavy-flavor jet classification. In these studies, various types of jet data representations and neural network architectures have been employed to make use of the information from jet constituents and various reconstructed substructures as much as possible, such as using the jet image with convolutional neural network (CNN), using the jet primary Lund plane with recurrent neural network (RNN), using the jet declustering history tree with recursive neural network (RecNN) and using four momentum information of jet components (point cloud) or jet Lund plane with graph neural network.

In relativistic heavy ion collisions, machine learning techniques are also applied in the study of jet quenching with the above motivations, including reconstructing the jet momentum~\cite{haake2019machine,Haake:2019pqd,Bossi:2020hwt,bossi2022r,ALICE:2023waz,Mengel:2023mnw}, distinguishing between quenched and unquenched jets~\cite{apolinario2021deep,Liu:2022hzd,lai2021information,romao2023jet}, identifying the jet energy loss~\cite{Du:2020pmp,du2021jet,du2021jeteps}, locating the jet creation points~\cite{yang2023deep} and classifying quark and gluon jets in the heavy-ion collisions~\cite{Chien:2018dfn,du2021classification}. In the following, we will review the applications of machine learning on these topics and give an outlook for the future studies.

\section{Reconstruction of jet momentum in heavy ion collisions}
In heavy ion collisions, the background fluctuations from the QGP medium pose huge challenges to the reconstruction of jet momentum. The conventional standard method for the "uncorrelated" background subtraction is area-based. The momentum of reconstructed jet is estimated by $p^{\mathrm{rec}}_{\mathrm{T,\, jet}}= p^{\mathrm{raw}}_{\mathrm{T,\, jet}} - \rho \cdot A$, which makes use of the event-averaged background density $\rho$ and jet area $A$. "Uncorrelated" here means that the effect of jet-medium interaction is not taken into account. This conventional method is based on an event-by-event analysis. The density of the low-energy background particles $\rho$ is derived from the median density of the clusters of low-energy particles other than the jet in the event of heavy ion collision. This background subtraction method brings large residual fluctuations, especially posing huge challenges for the momentum reconstruction of jets with $p_T<100$ GeV and large-radius. 

With machine learning methods, R. Haake and C. Loizides reconstruct the momentum of jets observed by the ALICE detector, where only charged particles can be observed at that time~\cite{haake2019machine}. The jet samples generated by the PYTHIA 6.4 model~\cite{sjostrand2006pythia} are embedded in the QGP background particles simulated by a simple thermal model, where the multiplicity distribution of the particles satisfies a realistic Gaussian distribution, and the momentum distribution is a modified power law distribution. The true value of the jet momentum after the background subtraction $p^{\mathrm{true}}_{\mathrm{T,\, ch\,jet}}$ is defined by considering the momentum fraction of the PYTHIA particles in the reconstructed jet,
\begin{equation}
\centering
  p^{\mathrm{true}}_{\mathrm{T,\, ch\,jet}} = p^{\mathrm{raw}}_{\mathrm{T,\, ch\,jet}} \cdot \sum_i p_{\mathrm{T,\;const}\;i}^\mathrm{PYTHIA}/ \sum_i p_{\mathrm{T,\;const}\;i}.
\end{equation}
Several machine learning methods, including shallow neural networks, random forests and linear regression, are employed to reconstruct the jet momentum with subtracting the QGP background $p^{\mathrm{rec}}_{\mathrm{T,\, ch\,jet}}$ in a supervised learning manner. Various jet observables are taken as inputs, including the uncorrected jet momentum, the jet momentum corrected by the area-based method, several jet shapes observables, the number of constituents within the jet, mean and median of all constituent transverse momenta, and the transverse momenta of the first 10 hardest particles within the jet. It is found that, since the machine learning methods consider both the density of background particles and the characteristics of the jet itself, i.e., a jet-by-jet analysis, their new results are superior to that of the established standard area-based method. It's verified that the trained neural network estimator is robust to thermal background of different multiplicities and anisotropies and allows for the application on the measurement of jets down to extremely low transverse momentum, e.g., 20 GeV, or of different jet resolution parameters. Besides, despite observing a small bias towards the jet fragmentation function, the ML-based methods are shown to be generalizable to the low $p_T$ jet samples simulated by the HIJING model~\cite{wang1991hijing}. Afterwards, the bias towards the fragmentation of PYTHIA jets have been investigated in more detail with several different modifications and the toy thermal model in~\cite{Bossi:2020hwt,ALICE:2023waz}, which demonstrates the robustness of the ML-based method to such possible bias. With the verification, the ML-based method has been applied to the experimental data in ALICE for charged jets and full jets~\cite{Haake:2019pqd,Bossi:2020hwt,bossi2022r,ALICE:2023waz}.  

To address the interpretability of the above ML-based method~\cite{haake2019machine}, T. Mengel~\textit{et al.} develop a multiplicity-based method as an alternative to the area-based method to subtract the background from jets in heavy ion collisions~\cite{Mengel:2023mnw}. In this method,
$p^{\mathrm{rec}}_{\mathrm{T,\, jet}}= p^{\mathrm{raw}}_{\mathrm{T,\, jet}} - \rho_{\mathrm{Mult}}\cdot(N_{\mathrm{tot}}-N_{\mathrm{signal}})$, where $N_{\mathrm{tot}}$ is the total number of particles in the jets,
$N_{\mathrm{signal}}$ is the number of particles in the signal other than the background and $\rho_{\mathrm{Mult}}$ is the mean transverse momentum per background particle in an event. This multiplicity method shows a lower variance of momentum residual than the area method and gives comparable performance with that of the neural network in Ref.~\cite{haake2019machine}. With the help of symbolic regression as an interpretable machine learning method, the authors argue that the trained neural network in Ref.~\cite{haake2019machine} is using a relationship similar to the multiplicity-based method. Furthermore, they emphasize the advantage of the interpretable machine learning methods in providing clear understanding of the methods especially when applied outside the range of the training data and estimating the measurement uncertainty.

In the above studies, the jet-medium interaction or the correlated background has not been considered. How to reasonably define the jet momentum in such a more realistic scenario and apply machine learning methods for its reconstruction should be further studied in the near future.

\section{Discriminating between quenched and unquenched jets}
In order to investigate the modification effects of jets in the QGP, several groups have carried out research to distinguish between unquenched jets and quenched jets based on MC simulations with different machine learning  methods~\cite{apolinario2021deep,Liu:2022hzd,lai2021information,romao2023jet}. In such a classification task, the labeling of the samples is straightforward and clear. In this section we will review these works.

To explore how different deep learning methods discriminate between the quenched and unquenched jets, L. Apolinário~\textit{et al.} use several neural network architectures and jet data representations to classify the unquenched and quenched jets~\cite{apolinario2021deep} with the $Z$ + jet samples generated by the Monte Carlo generator JEWEL 2.0.0~\cite{zapp2014jewel}. The underlying events and recoil scattering are ignored in their study. They use CNN, RNN and fully-connected neural networks with jet image, jet primary sequence of Lund plane and two jet observables (jet transverse momentum and number of jet constituents), respectively, as inputs for the classification task. With these setups, they achieve the best classification performance with CNN taking the unnormalized jet image as the input and the second best with RNN taking Lund plane as the input. Different jet data representations carry different physical information. This comparison helps us understand how deep learning identifies the type of jets. Besides, since the $Z$ + jet samples are considered, the ratio of the transverse momentum between the jet and the $Z$ boson $x_{jZ}=p_{T,jet}/p_{T,Z}$ can serve as an effective measure of the jet quenching degree. 
The correlations between the outputs of these neural networks and $x_{jZ}$ are examined to conclude that these neural networks can effectively serve as the classifiers of jet quenching.

In Ref.~\cite{Liu:2022hzd}, L. Liu~\textit{et al.} use a long short-term memory (LSTM) neural network to classify the unquenched jets generated by PYTHIA 8 and the quenched jets generated by JEWEL. They take the primary Lund sequence of jets as inputs for this classification task. To take into account the effect of uncorrelated thermal background for the quenched jets, underlying events are embedded properly in both event generators. 
To tackle the non-determinism of the raw outputs with respect to different trained LSTM models, the authors design a calibration method accordingly. With the identification of quenched jets, the modifications of jet substructures by the quenching effect are investigated by looking at the top 40\% and bottom 60\% quenched jets, respectively. The quenched class (top 40\%) shows obvious modifications of jet substructures with the enhancement of wider and softer splitting. While the less quenched class (bottom 60\%) shows a similar quenching pattern with the unquenched jets. 

To identify the important features with high discriminating power for the quenching effects, Y. S. Lai~\textit{et al.} classify the unquenched jets generated by PYTHIA 8~\cite{Sjostrand:2007gs} and the quenched jets generated by JEWEL 2.2.0 with the Particle Flow Network (PFN)~\cite{komiske2019energy} and Energy Flow Network (EFN)~\cite{komiske2019energy}, taking IRC-unsafe and IRC-safe jet observables as inputs for this classification task~\cite{lai2021information}. It is found that the PFN gives better classification performance, which shows that a large amount of jet quenching information is contained in IRC-unsafe physics. Besides, dense neural network (DNN) and linear classifier are used for this classification task, taking complete sets of IRC-safe observables, i.e., N-subjettiness~\cite{thaler2011identifying,thaler2012maximizing} and Energy Flow Polynomials (EFP)~\cite{komiske2018energy} as inputs. Through a detailed comparative study with different number of observables, it's demonstrated that a large amount of jet quenching information is contained in the soft radiations within the jets, which is consistent with the observation in Ref.~\cite{Du:2020pmp}. They also design the novel analytical observables which are highly-discriminating and understandable. In addition, the changes in classification accuracy when including and then subtracting the background in heavy ion collision are presented. It is found that including the background will significantly reduce the classification accuracy while subtracting the background will slightly reduce the accuracy further. On the one hand, the verification shows that part of the jet quenching information is irreversibly lost in the presence of the QGP background, which poses a challenge for the further applications of deep learning techniques to the realistic experimental data. On the other hand, the difference between the classification performance before and after the background subtraction can serve as a measure of the capability of different background subtraction schemes to retain relevant information and help us select and optimize the background subtraction schemes.

In Ref.~\cite{romao2023jet}, M. C. Romão~\textit{et al.} perform three analyses to study the correlations between jet substructures for quenched and unquenched jet samples using JEWEL+PYTHIA, respectively, and check their robustness to the quenching effects. Firstly, the authors employ the Principal Component Analysis (PCA) method in an unsupervised manner to study the linear correlations between 31 jet substructures and find that the first 10 principal components can explain 90\% of the distributions of all observables. Secondly, they use the AutoEncoder method also in an unsupervised manner to capture the non-linear correlations between these observables and find that 10 latent degrees of freedom can encode almost full information while 5 ones are enough to capture above 90\% information. Besides, compared with principal components, same number of latent degrees of freedom learned in AutoEncoder can give systematically better quality of jet reconstruction, which highlights the importance of non-linear correlations between jet observables. Thirdly, they use Boosted Decision Trees (BDT) method to check the discriminating power of specific and pairs of observables between unquenched and quenched jets compared with the full set. The identified observables can serve as the optimal candidates for taggers of quenching effects. Last but not least, it is found that, though the robustness of correlations among jet substructures to the quenching effects, quenching effects manifest themselves by the strong population migration of these observables.

\section{Identifying jet quenching degree}
Compared to unquenched jets, quenched jets actually have a larger diversity, which is mainly due to the fact that the degree of jet modifications are affected by the in-medium traversed length of jets and the temperature of medium along the propagation of jets. In this sense, unquenched jets can be viewed as the quenched jets with vanishing in-medium traversed length. Therefore, though it is worth trying to find a classifier that separates these two types of jets maximally, the physical interpretation on the outputs of the trained neural networks is unclear, which hinders their applications on the experimental data. Compared with the task of distinguishing between quenched jets and unquenched jets, switching to identify the degree of in-medium modifications of jets could be more experimentally applicable in the future.

We use CNN to predict the jet energy loss on a jet-by-jet basis with the jet image in a supervised manner~\cite{Du:2020pmp,du2021jet}. 
A hybrid strong/weak coupling model~\cite{casalderrey2015erratum,Casalderrey-Solana:2015vaa} is used to simulate the energy loss process of jets in the QGP created in Pb-Pb collisions. 
The amount of energy loss, quantified through the ratio variable $\chi \equiv \pTf/\pTi$ suffered by jets due to the propagation through a hot and dense QCD medium. $\pTf$ is the transverse momentum of a given jet in the presence of a medium with cone size $R$, and $\pTi$ is the transverse momentum of the \emph{same} jet had there been no medium, see \cite{Du:2020pmp} for further details on how to establish such a correspondence. A good prediction accuracy on the energy loss ratio $\chi$ is achieved after the training and validating of the neural network.

From the average jet images normalized by jet $p_T$ sliced in several different $\chi$ bins, one can clearly see that as the jet energy loss increases, more soft particles gradually populate at large angle within the jet, which provides an intuitive understanding of the success for our prediction task. In addition to using jet images as input, we also use some jet observables and their combinations as the inputs to fully-connected neural networks to perform the same task in parallel for comparison and interpretation~\cite{Du:2020pmp}. These jet observables include jet fragmentation function (JFF), jet shape, and some single-value jet observables, such as jet $p_T$, jet mass $M$, jet multiplicity and some groomed jet substructures $z_g$, $n_{SD}$, $R_g$, $M_g$. It is found that the performance given by the jet fragmentation function, jet shape and jet features increase. If combining jet fragmentation function and jet shape as inputs, their performance will be closer to the performance given by the jet image normalized by $p_T$. If all these observables are used as inputs, they can reproduce the performance given by the jet image. This can provide an indirect interpretation for the success of CNN prediction using the jet image as the input. 

With the prediction of jet energy loss at hand, many interesting applications are allowed to make. Usually one selects jets according to their final, measured energy (FES), such that they are above a certain momentum threshold, $p_T>p_T^{\mathrm{cut}}$. By estimating the energy lost by the jets in the medium, one can have an initial energy selection (IES) instead, i.e., $\pTi>p_T^{\mathrm{cut ,initial}}$ (at the same time we also require $p_T>p_T^{\mathrm{cut}}$ to ensure that it is within the scope where the neural network is trained. As long as $p_T^{\mathrm{cut , initial}}$ is sufficiently higher than $p_T^{\mathrm{cut}}$, we are actually considering all the jets of $\pTi>p_T^{\mathrm{cut,initial}}$). We show that this novel selection method can effectively remove the selection bias induced by the final-state interaction of the jets with the QGP, which helps us to reveal the quenching effects on jet substructures, e.g., $z_g$, $n_{SD}$ and $R_g$~\cite{Du:2020pmp}, as well as get access to the genuine spatial distribution and the possible initial-state anisotropy of the jets over the transverse plane of the nuclear collision~\cite{du2021jet}.

\section{Jet tomographic study of quark gluon plasma}

As aforementioned, jet energy loss is strongly related to the traversed length of jets in the QGP. That is to say, selecting the jet samples that lose different degrees of energy is actually selecting the jet samples created at different positions of the QGP. We demonstrate this picture within the strong/weak hybrid coupling model where the creation positions of jets are taken from the Monte Carlo simulations directly.
By looking at the creation-point distributions of the jets in the transverse plane of collisions sliced in different energy loss ratio $\chi$ bins, one can find that the jets that lose little energy is mainly distributed in the surface of the QGP. As the energy loss increases, the creation positions of jets are gradually distributed towards the central area of the QGP. It is worth noting that for the jets that lose a lot of energy, their creation positions will be away from the central area again, and they will pass through the central area in the opposite direction to obtain a longer traversed distance or traverse the areas with higher temperature~\cite{Du:2020pmp}. This picture can be seen more clearly if one only considers the jets with certain direction~\cite{du2021jet}. Besides, constraining the measured jet substructures additionally will allow us to have a finer selection of the creation positions of jet samples~\cite{Du:2020pmp,du2021jeteps}. The above observation serves as an important step forward towards the jet tomographic study of QGP.

In Ref.~\cite{yang2023deep}, Z. Yang~\textit{et al.} employ a point cloud neural network to locate the initial jet production positions directly with the full information of the final hadrons inside the jets and the global information of $\gamma$ and jet in $\gamma$-jet events in heavy-ion collisions. The training of the neural network is done on the jet samples from the CoLBT-hydro model~\cite{chen2018effects} and the generalizability is examined on the jet samples from the LIDO model~\cite{ke2019modified}. The validating and testing errors with the form of root mean square are around 2.2$\sim$2.4 fm, respectively. With the prediction of initial jet production locations, the signal of Mach cone and diffusion wake in the expanding QGP can be amplified by selecting jet samples within specific spatial regions to have long jet traversed lengths with the known propagation direction relative to the radial flow. This deep learning assisted jet tomography will help confirm the existence of Mach cones in experimental data. Besides, one can use this method to investigate the path length dependence of jet energy loss form the measured $\gamma$-jet $p_T$ distributions.

\section{Discriminating between quark and gluon jets in heavy ion collisions}

Quark- and gluon-initiated jets could become independent probes since their partonic origins experience different evolution, both in vacuum and medium. In vacuum, their microscopic processes and corresponding splitting functions are different. The splitting angle of gluon are larger on average, which indicates a larger phase space for the medium-induced radiative energy loss. In addition, the in-medium traversed length dependence of quark and gluon partonic energy loss is different with a color factor. Therefore, with the same traversed length, the gluon jets will lose more energy on average than quark jets and they will be modified by the medium differently.

To use quark and gluon jets as independent probes to the QGP, it would be interesting to explore the differences between these two types of jets and classify them~\cite{Chien:2018dfn,brewer2021data}. If one can select purer samples of quark and gluon jets, it will help us pin down the dependence of energy loss of jets with different flavor on the in-medium traversed length, and push forward the jet tomographic study. Another reason for distinguishing between quark and gluon jets is to check the universality of energy loss for different processes, e.g. comparing dijet events (with a mix of quark- and gluon-initiated jets) with boson-jets (where the parton recoiling from the boson is predominantly a quark).

In Ref.~\cite{Chien:2018dfn}, Y.-T. Chien and R. K. Elayavall use physics-motivated jet observables, jet image and jet substructures defined by the telescoping deconstruction (TD) method, respectively, as inputs to identify the quark and gluon jets in the pp and Pb-Pb collisions generated by JEWEL. Both CNN and the TD method can give the best classification performance. Furthermore, it is found that for the quenched jets in heavy ion collisions, the discriminating performance of the TD method will decrease due to the significantly increased soft particles that affect the jet substructures. 

In Ref.~\cite{du2021classification}, we use deep learning techniques to perform a similar task with the strong/weak hybrid model. By looking at the average of $p_T$-normalized quark and gluon jet images sliced in different energy loss ratio $\chi$ bins, one can find that with the increase of energy loss, quark jets and gluon jets show the same qualitative characteristics, i.e., there are more soft particles populating at large angle in the jet cone. But the quantitative characteristics is different. Losing the same fraction of energy, gluon jets look wider than quark jets, and there are more soft particles at large angle area. In fact, the gluon jets losing a certain amount of energy, i.e. at a given $\chi_g$ look like quark jets that lost significantly more energy, i.e. with a $\chi_q \ll \chi_g$. 
We employ a CNN to classify the quark and gluon jets from jet images. It is found that, overall, the accuracy of classifying quark and gluon jets in the medium is a bit lower than that in the vacuum, and the decrease of accuracy depends on the fraction of the quenched jets. It can also be inferred from the results of classification for quenched jets with different $\chi$, i.e., the greater the energy loss is, the more difficult the classification is. This observation agrees with that in Ref.~\cite{Chien:2018dfn}.

\section{Summary and Outlook}
In this proceedings, we have comprehensively reviewed the applications of machine learning techniques in jet quenching physics, including the background subtraction for jet momentum reconstruction in heavy ion collisions, distinguishing between quenched jets and unquenched jets, determining the degree of jet energy loss and the initial jet creation positions in QGP and classifying quark jets and gluon jets in heavy ion collisions. On the bases of these studies, many interesting applications have been made, including revealing the genuine modifications of jet substructures due to quenching effects, getting access to the spatial and angular distribution of the initial jets production in the transverse plane of collisions and amplifying the signal of Mach cone and diffusion wake induced by the jets in the expanding QGP. 

In the future studies, it is very essential to embed the correlated fluctuating background in the simulated jet samples to mimic the realistic scenario and examine the generalizability of some applications between different jet quenching Monte Carlo models, which are crucial to their final applications to the experimental data. In addition, many interesting applications deserve to be explored in the near future, such as directly determining the traversed length by the jets in the QGP to pin down the path-length dependence of jet energy loss, reconstructing the substructures of the twin vacuum partner of the quenched jets to build up the one-to-one correspondence in the aspects other than the transverse momentum. In these tasks, novel types of neural network architectures associated with jet data representations and quantum machine learning method are worth exploring to improve the performance and get better generalizability.



\acknowledgments
This work is supported by the Taishan Scholars Program and Shandong Excellent Young Scientists Fund Program (Overseas) (No. 2023HWYQ-106).

\bibliographystyle{apsrev4-1}
\bibliography{duyl}

\end{document}